\documentclass[
 amsmath,amssymb,
 prd,
 notitlepage,
]{revtex4-1}

\usepackage{graphicx}
\usepackage{dcolumn}

\usepackage{natbib}

\usepackage{bm}
\usepackage{xcolor}
\usepackage{mathrsfs}
\usepackage{hyperref}

\begin{document}

\title{Trapped region in Kerr-Vaidya space-time}

\author{Pravin Kumar Dahal}
 \email{pravin-kumar.dahal@hdr.mq.edu.au}

\affiliation{
 Department of Physics and Astronomy, Macquarie University.
}
\date{\today}

\begin{abstract}
    We review the basic definitions and properties of trapped surfaces and discuss them in the context of Kerr-Vaidya line-element. Our study shows that the apparent horizon does not exist in general for axisymmetric space-times. The reason being the surface at which the null tangent vectors are geodesic and the surface at which the expansion of such vectors vanishes do not coincide. Calculation of an approximate apparent horizon for space-times that ensure its existence seems to be the only way to get away with this problem. The approximate apparent horizon, however, turned out to be non-unique. The choice of the shear free null geodesics, at least in the leading order, seem to remove this non-uniqueness. We also propose a new definition of the black hole boundary.
\end{abstract}

\maketitle

\section{Introduction}

The notion of the trapped surface was first introduced by Penrose\cite{31}, and the concept later emerged as a quasi-local technique to characterize a black hole. They are seen as a replacement to the event horizon to define the black hole boundary, which depends on the global knowledge of the space-time. The teleological nature\cite[]{32,19} and the unobservability\cite[]{30} of the event horizon have been discussed numerously in the past.  It has thus been regarded as unsuitable for characterizing astrophysical black holes, which are dynamical in nature and has to be studied by a finite-sized laboratory for the finite time. However, even though the trapped surfaces are quasi-local making them observable in principle, they are not a well-defined quantity to characterize black holes. The principal reason behind this is the foliation dependence of the trapped surface. As the choice of the foliation depends on the choice of null vectors, the trapped surface depends on the choice of null vectors\cite[]{33}.

Defining a black hole boundary has been like a reverse engineering process. We somehow know the answer for some simple cases based on which we want to define a black hole boundary such that the definition reproduces that answer. Such simple cases are stationary space-times of spherical and axial symmetry for which all the horizons (event, apparent, killing, trapping horizons) coincide. The notion of trapped surfaces, introduced in this way to define a black hole boundary poses no problem at all for stationary space-times, even though trapped surfaces depend on the choice of foliations for these space-times\cite[]{18}. Then comes the second stage of making the use of the definition of the trapped surface to identify the black hole boundary of the space-time for which the concept of an event/ killing horizon fails down. Trapped surface as a black hole boundary has celebrated its success in application to the general spherically symmetric space-times. The problem of foliation dependence, in that case, can be eliminated by making the particular choice of foliation that respects the symmetry of space-time\cite[]{33}. All spherically symmetric foliations give the same trapped surface for such space-times. This problem, however, is more visible and profound if we move to the more general case of axially symmetric space-times, and there do not exist simple solutions of this problem in axial symmetry.

This article is the introduction of the trapped surface and its application to the axially symmetric space-time by taking the reference of the Kerr-Vaidya metric. We introduce trapped surface, define various types of trapped surfaces, and discuss, in brief, the notion of the geometric horizon as a black hole boundary in Sec.\ref{1c}. Recently, the geometric horizon has been considered as an alternative over the trapped surface in locating the black hole boundary because of its invariant characterization, quasi-local nature, and foliation independence. However, its applicability has been demonstrated only for limited classes of space-times until now. In Sec.\ref{1d}, we present some of the properties of the trapped surfaces based on their metric, and in Sec.\ref{s2e}, we introduce the slowly evolving horizon. The concept of a slowly evolving horizon is useful in the sense that it does not mingle with the problem of foliation dependence, provided that the condition for its existence satisfies. Sec.\ref{2} is the application of the trapped surface to the axisymmetric space-time, and for this, we chose one of the simplest example of Kerr-Vaidya geometry in advanced coordinates. We also discuss our results in the context of the geometric horizon. Finally, in Sec.\ref{4}, we propose a new black hole boundary.

\subsection{Trapped surfaces and Geometrical horizons}\label{1c}

We begin here with the notion of an event horizon, which is believed to be the true boundary of a black hole by some authors despite its teleological nature.

Asymptotically flat space-time containing black holes can be separated into two regions: 1) the region from which all null curves can reach the future null infinity (where future-directed null curves end in an infinite time in the Minkowskian space-time). 2) The region from which no null curve reaches the future null infinity. A boundary separating these two regions of space-time is defined as an event horizon. This definition of the event horizon depends on the complete knowledge of future-directed null curves, which is impossible to achieve in a finite time making it physically unobservable\cite[]{30}. This global nature of the event horizon also makes it a less useful entity for dynamical space-times. Similarly, the definition of the event horizon could not be applied to the space-times which are not asymptotically flat. Thus, the notion of the trapped surfaces that depends on the quasi-local nature of space-time is handy in describing black holes evolving in time. In principle, it is not possible to define a horizon that can be detected by completely local measurements. The reason behind this is rooted in the equivalence principle, which implies an impossibility of measuring space-time curvature by local measurements.

In the orientable space-time manifold ${\cal M}$, lets take future-directed field of null vectors $l^\alpha$ and $n^\alpha$ satisfying $l^\alpha n_\alpha=-1$. We assume that $l^\alpha$ is outgoing in the sense it is pointed away from the trapped region, and $n^\alpha$ is ingoing. The divergence (or the convergence) of the field of null congruence $l^\mu$ is given by its covariant derivative $l^\mu_{;\mu}$. If the vector field $l^\mu$ is arbitrarily parametrized, then we affinely parametrize it before the calculation of divergence, and this introduces the additional correction term $\kappa=-n_\mu l^\nu l^\mu_{;\nu}$, called the surface gravity. Thus, the divergence (or convergence) of the affinely parametrized null congruence is given as\cite[]{2}
\begin{equation}
    \theta_l=l^\mu_{;\mu}+n_\mu l^\nu l^\mu_{;\nu} \qquad \textrm{and} \qquad  \theta_n=n^\mu_{;\mu}+n_\mu l^\nu n^\mu_{;\nu},
    \label{e1}
\end{equation}
where $\theta_l$ denotes the outgoing expansion, and $\theta_n$ denotes the ingoing expansion (for details on the expansion of congruence see textbook by Poisson\cite{3}). Among these, the condition for the vanishing of outgoing expansion $\theta_l=0$ is pivotal for the definition of trapped surfaces. A surface is said to be trapped if $\theta_l<0$ and $\theta_n<0$. Thus on the trapped surface, both the ingoing and outgoing null geodesics are converging. A marginally outer trapped surface (MOTS) has $\theta_l=0$ and $\theta_n<0$. This surface outlines the boundary from which the outgoing null geodesics start to converge. The three-dimensional surface which can be foliated entirely by the MOTS is called a marginally outer trapped tube (MOTT). This MOTT is defined as the future apparent horizon in recent literature\cite[]{27,12}, and we will adopt this definition here. The definition of an apparent horizon and, in general, of the trapped surfaces varies over the literature. The Hayward's trapping horizon is defined as an apparent horizon satisfying $n^\mu (\theta_l)_{;\mu} \ne0$\cite[]{29}. This additional condition imposed implies that the space-time surface is trapped inside the future outer trapping horizon (FOTH) and normal outside it. Similarly, a spacelike apparent horizon is defined as a dynamical horizon\cite[]{32}.

The definitions of trapped surfaces given here provide an easy way of identifying black holes and these are just by the calculation of null expansions $\theta_l$ and $\theta_n$. We thus do not have to wait till the end of time to know the formation of an event horizon to identify black holes. Trapped surfaces indeed are convenient and unique for the numerical and analytical study of spherically symmetric space-times (for spherically symmetric foliation, which is an obvious choice of foliation because of the symmetry of space-time). However, even the trapped surfaces not being entirely local entities, their identification depends on the closed hyper-surface $\Sigma$ of the space-time whose foliation spans the whole space-time. Expansions should be calculated over all the points of the closed hyper-surface $\Sigma$, not just at a single point for their identification. Hence, trapped surfaces are quasi-local in nature and depend on the complete knowledge of the hypersurface $\Sigma$, if not of the space-time.

As mentioned above, with any choice of spherically symmetric foliations, trapped surface in spherically symmetric space-time is unique. However, the complication starts to appear when we consider foliations that are not spherically symmetric, even to calculate the trapped surface in spherically symmetric space-times. The foliation dependence of trapped surface in Vaidya space-time has been demonstrated by numerical calculations with some special choice of non-spherically symmetric foliation in Refs.\cite[]{16, 17}. Discussion of the foliation dependence of the trapped surface in Schwarzschild and Vaidya space-time can also be found in the review by Krishnan\cite{18}. We will describe the trapped surface in Kerr-Vaidya space-time and its foliation dependence.

Another type of horizon that is commonly used to characterize the black hole boundary is the notion of geometric horizon\cite[]{15g,16g}. The existence of the geometric horizon is assured by the conjecture that the non-stationary black holes contain the hypersurface which is more algebraically special\cite[]{14g,15g} and such a hypersurface is identified as the geometrical horizon. This method of locating the black hole does not suffer the problem of foliation dependence. 
The property of being more special can be quantified invariantly by making particular combinations of curvature invariants zero on the horizon. Three different sets of invariants derived from curvature tensors are commonly used in the literature to calculate the geometric horizon in general. The first approach relies on finding the suitable Killing vector field that becomes the null generator on the horizon. Then, the hypersurface where the square norm of the Killing vector vanishes is the event horizon. If it is not possible to find such a field of Killing vector, then scalar polynomial curvature invariants could be calculated for stationary space-times and space-times conformal to stationary space-times\cite[]{21g}. At the event horizon of such space-times, it has been shown that the squared norm of the wedge products of the n-linearly independent gradients of scalar polynomial curvature invariants vanishes, n being the local cohomogenity of space-time\cite[]{17g,18g}. This method of locating the black hole boundary is restricted to the stationary space-times as this method, in principle, is looking for the null surface where the magnitude of some particular combinations of invariants vanishes. However, the horizons of dynamical space-times are not null in general. The extension of this method to the dynamic space-times conformal to the stationary space-times lies on the fact that the event horizon is invariant under conformal transformation.

Another procedure for calculating the geometric horizon is by finding the zeroes of the certain combinations of Cartan invariants\cite[]{19g,20g}. This method is exactly similar, in principle, to the scalar polynomial invariants for finding the geometrical horizon. Thus, this method is also limited to stationary space-times and any space-times conformal to them. However, Cartan invariants are regarded as the improvement over the scalar polynomial invariants as this method involves the linear combinations of the components of curvature tensor, and they are possible to construct from Cartan invariants.

The third procedure for the calculation of the geometric horizon is similar to the calculation of the trapped surfaces, and thus the black hole boundary calculated using this procedure can be called as geometrically trapped surface. The calculation is based on the conjecture that the hypersurface constituting geometrically trapped surface is more algebraically special\cite[]{17g,18g}. The procedure is to look for the geometrically preferred outgoing null vectors and then find the hypersurface where the expansion of such vectors vanishes. The null vector is called the geometrically preferred null vector because the congruence of the null vectors thus chosen is such that the covariant derivatives of the curvature tensor are more algebraically special there\cite[]{14g,16g}. The applicability of this method extends beyond the above two procedures of finding the geometric horizon as this method can be extended to the dynamical space-times also. However, until now, it has not been used to locate the black hole boundary of any axisymmetric dynamical space-times. The equivalence/correspondence between these three procedures for calculating the geometrical horizon lies in the fact that they all involve the calculation of invariants derived from the curvature tensor (see, for example, \cite{14g}, \cite{20g} for details).

\subsection{Properties of trapped surfaces} \label{1d}

Refs.\cite{22,23} presented a simple formalism to determine the metric signature of an apparent horizon and demonstrated that for spherically symmetric space-times. We will here discuss it for the purpose to determine some properties of the MOTT of the Kerr-Vaidya geometry. Let, $\chi^\mu$ be the vector field tangential to the MOTT and orthogonal to the MOTSs that foliate the MOTT. Also assume that the vector field $\chi^\mu$ generates the flow that preserves the foliation. Then, ${\cal L}_\chi v=f(v)$, where ${\cal L}_\chi$ denotes the Lie derivative along $\chi$ and $f(v)$ denotes some function on the foliation $v$. We similarly denote by $\bar{\chi}^\mu$ the vector field normal to the MOTT which satisfies $\bar{\chi}_\mu \chi^\mu =0$. If we assume both $\chi^\mu$ and $\bar{\chi}^\mu$ are future directed, then one of them is spacelike and another is timelike. Now, in terms of the null vectors $l^\mu$ and $n^\mu$ given above, we can write:
\begin{align}
    & \chi^\mu=l^\mu-C n^\mu,\\
    & \bar{\chi}^\mu=l^\mu+C n^\mu,
\end{align}
for some function $C$. Along $\chi^\mu$, the Lie derivative of an outgoing expansion is zero thereby implying
\begin{equation}
    {\cal L}_\chi \theta_l={\cal L}_l \theta_l-C {\cal L}_n \theta_l=0,
\end{equation}
or,
\begin{equation}
    C=\frac{{\cal L}_l \theta_l}{{\cal L}_n \theta_l}.
    \label{e16}
\end{equation}
Eq.\eqref{e16} can be used to calculate $C$ and as $C\propto \chi^\mu \chi_\mu$, the metric signature of the MOTT is determined by the sign of $C$. $C>0$ means that an apparent horizon is spacelike, $C<0$ means that an apparent horizon is timelike and $C=0$ or $\infty$ means that an apparent horizon is null. The sign of $C$ also determines whether an apparent horizon is expanding or contracting. To see this, lets take an area element of the two-surface be $\tilde{q}$ and evaluate its Lie derivative along $\chi^\mu$:
\begin{equation}
    {\cal L}_\chi \tilde{q}=-C \theta_n \tilde{q}
    \label{e17}
\end{equation}
As $\theta_n<0$ on trapped surface, $C>0$ implies that an apparent horizon is expanding, $C<0$ implies that the horizon is receding and $C=0$ or $\infty$ implies that the horizon is isolated. Thus, the spacelike apparent horizon is expanding and the timelike apparent horizon is receding, while an isolated apparent horizons are null.

\subsection{Slowly evolving horizon}\label{s2e}

We are working in the semi-classical domain for which the quasi-static evaporation law given in \cite{26,27} is valid. In this domain, an apparent horizon of the Kerr-Vaidya metric can be approximated as a slowly evolving horizon. So, we present some characteristics of the slowly evolving horizon here. An almost-isolated trapping horizon is defined as a slowly evolving horizon and the invariant characterizations of such horizon are made by Booth and Fairhurst\cite{24,25}. We have mentioned above that an apparent horizon is null and isolated if $C=0$. It thus seems that a slowly evolving horizon can be characterized by the spacelike or timelike horizon for which $C$ is small. $C$, however, can be arbitrarily varied by rescaling $n^\mu\to \frac{n^\mu} {\alpha}$ and $C\to \frac{C}{\alpha^2}$, for some $\alpha$. The slowly evolving horizon should thus be identified in a scaling independent manner and for this, we write the area evolution law of Eq.\eqref{e17} as:
\begin{equation}
    {\cal L}_\chi \tilde{q}=-||\chi|| \bigg(\sqrt{\frac{C}{2}}\theta_n \tilde{q}\bigg)
\end{equation}
such that, the term in parentheses is independent of rescaling. This term gives an invariant area evolution and if this term is small, then the horizon is slowly evolving. Thus, one of the conditions proposed by Booth and Fairhurst\cite{24} for the horizon to be slowly evolving over some MOTS in a given foliation is
\begin{equation}
    \sqrt{C}\theta_n<\frac{\epsilon}{R},
\end{equation}
for arbitrary small $\epsilon$ where $R$ is the areal radius of the MOTS. Another condition is the choice of scaling of the null vectors, such that, $||\chi||\sim\epsilon$. There are other conditions given in \cite{24,25} requiring that the horizon evolves smoothly over a long period of time. These conditions will be automatically satisfied for the slowly evolving Kerr-Vaidya metric where the horizon offsets from the isolated horizon, at most, by an order of $M_v$. The reason behind this is, every parameter calculated at the horizon, in this case, differs from the stationary Kerr metric by an order of $M_v$.

\section{Trapped region in Advanced Kerr-Vaidya space-time} \label{2}

\subsection{Some Examples}\label{2a}

We have, the line-element for Kerr-Vaidya space-time in advanced coordinates $(v, r, \theta, \tilde\phi)$ is given as\cite[]{1}:
\begin{equation}
    ds^2=-\bigg(1-\frac{2 M(v) r}{\rho^2}\bigg)dv^2+2 dv dr+\rho^2 d\theta^2-
    \frac{4 a M(v) r \sin^2\theta}{\rho^2}d\tilde\phi dv
    -2 a \sin^2\theta d\tilde\phi dr+
    \frac{(r^2+a^2)^2-a^2 \Delta \sin^2\theta}{\rho^2}\sin^2\theta d\tilde\phi^2,
    \label{1}
\end{equation}
where $\rho^2=r^2+a^2 \cos^2\theta$ and $\Delta=r^2-2 M(v) r+a^2$. This metric reduces to the familiar Kerr metric for some advanced time $v$ if $M(v)$ is constant\cite[]{6}. The trapped region associated with this metric can be determined by the calculation of the expansion scalars $\theta_l$ and $\theta_n$. To find the trapped surface for this metric, we use the future-directed null vectors of the form\cite[]{1}
\begin{equation}
    \begin{split}
        l_\mu=\frac{\rho^2}{2\Omega^2}\bigg(-\Delta,\; r^2+a^2+\Omega,\; 0,\; 0\bigg),\\
        n_\mu=\bigg(-1,\; \frac{r^2+a^2-\Omega}{\Delta},\; 0,\; 0 \bigg),
        \label{s9}
    \end{split}
\end{equation}
where $\Omega=\sqrt{(r^2+a^2)\rho^2+2 M(v) r a^2 \sin^2\theta}$. These null vectors also being normalized, satisfy the relations
\begin{equation}
    l_\mu l^\mu=n_\mu n^\mu=0 \; \text{and} \; l_\mu n^\mu=-1.
    \label{3}
\end{equation}
Moreover, calculation of the geodesic equation $l^\mu l_{\nu; \mu}=\lambda l_\nu$ for some parameter $\lambda$ shows that the vector $l^\mu$ is tangent to the null geodesic only on $\Delta=0$ surface. We now use Eq.\eqref{e1} to calculate the expansion scalars using these null vectors. The expansions thus obtained are given as
\begin{equation}
    \begin{split}
        \theta_l=\frac{1}{2 \Omega^3}\bigg(
        \Delta (2 r^3 + 2 a^2 r - a^2 r \sin^2\theta + a^2 M \sin^2\theta )+a^2 r M_v \sin^2\theta (r^2+a^2+ \Omega)\bigg),\\
        \theta_n=-\frac{1}{2 \rho^2 \Delta \Omega}\bigg(
        \Delta (2 r^3 + 2 a^2 r - a^2 r \sin^2\theta + a^2 M \sin^2\theta )+a^2 r M_v \sin^2\theta (r^2+a^2- \Omega)\bigg),
        \label{5}
    \end{split}
\end{equation}
where $M_v=\frac{\partial{M}}{\partial{v}}$. Now, a MOTS is the two-surface where outgoing null expansion ($\theta_l$) vanishes and this marks the boundary of the trapped region. Senovilla and Torres\cite{1} have argued that the solution for $\theta_l=0$ does not exist in general for the $v=constant$ foliation considered. This is indeed true and the reason behind is that the null vectors are geodesic only on $\Delta=0$ surface and on this surface, $\theta_l=0$ only when $M_v=0$. They have thus considered the surface of intersection between $\Delta=0$ and $M_v=0$ where $r_g=M\pm \sqrt{M^2 -a^2}$ are the two real solutions of $\theta_l=0$. For dynamical space-time considered, $M_v=0$ is not true in general and might occur only for some $v=v_0$. Thus, $v=constant$ surface does not foliate an apparent horizon in general.

Solution of $\theta_l=0$ is simpler and more evident if we calculate the expansion scalars for the null geodesics with the following tangent vectors:
\begin{equation}
    \begin{split}
        & l_\mu=\frac{\Delta}{2\rho^2}\bigg(-1,\; \frac{2\rho^2}{\Delta},\; 0,\; a \sin^2\theta \bigg),\\
        & n_\mu=\bigg(-1,\; 0,\; 0,\; a \sin^2\theta \bigg).
        \label{o8}
    \end{split}
\end{equation}
$l_\mu$ satisfy the geodesic equation $l^\mu l_{\nu; \mu}=\lambda l_\nu$, for some parameter $\lambda$ defining the geodesic curve only when $M_v=0$. The expansions using these pairs of null vectors are:
\begin{equation}
    \theta_l=\frac{r \Delta}{\rho^4}, \;\;\;\;\; \theta_n=-\frac{2 r}{\rho^2}.
    \label{o18}
\end{equation}
We thus get $\theta_l=0$ when $\Delta=0$. Hence, the MOTS should be the intersection between $\Delta=0$ and $M_v=0$ leading us again to the conclusion that $v=constant$ surface does not foliate an apparent horizon in general.

\subsection{MOTSs are not unique}\label{2b}

Here, we will first show that depending on the choice of the null vectors, the MOTS (the surface for which $\theta_l=0$) can extend all the way from the singularity to the infinity. For this, we take the general null vector of the form:
\begin{equation}
    \begin{split}
        l_\mu=(-\alpha, \; 1, \; \beta, \; \gamma)\\
        n_\mu=(-1, \;0, \; 0, \; a \sin^2\theta)
    \end{split}
\end{equation}
where $\alpha=\alpha(v,r,\theta, M(v))$. The outgoing vector $l_\mu$ will be future directed for $\alpha>0$ and it is null implies that $l_\mu l^\mu =0$. This allows us to express $\gamma$ in terms of $\alpha$ and $\beta$ as:
\begin{equation}
    \gamma= (-a+\alpha a \pm \sqrt{(1-2\alpha) a^2- (\Delta-2 \alpha (a^2+r^2)+\beta^2)\csc^2\theta})\sin^2\theta
\end{equation}
With this value of $\gamma$, the vectors $l_\mu$ and $n_\mu$ satisfies all the relations given in Eq.\eqref{3}. We can now calculate the outgoing expansion $\theta_l$ and its value is obtained as:
\begin{equation}
    \theta_l=\theta_{Kerr}+ \frac{2 a M_v \bigg(r+\rho^2\frac{\partial\alpha}{\partial M}-\beta\frac{\partial \beta}{\partial M}\bigg)}{\rho^2 \sqrt{(1-2\alpha) a^2- (\Delta-2 \alpha (a^2+r^2)+\beta^2)\csc^2\theta}}
\end{equation}
where $\theta_{Kerr}$ is an expression containing terms that does not have $M_v$. This corresponds to the outgoing expansion for the Kerr metric with $M=constant$. But, for the Kerr geometry, we should have
\begin{equation}
    \theta_{Kerr}= \frac{\Delta g(t, r, \theta)}{\rho^2 \sqrt{(1-2\alpha) a^2- (\Delta-2 \alpha (a^2+r^2)+\beta^2)\csc^2\theta}},
\end{equation}
 where $g(t, r, \theta)$ is an arbitrary function. However, we should have $g(t, r, \theta)>0$ so as to ensure that $\theta_{Kerr}>0$ for $\Delta>0$. We thus get
 \begin{equation}
     \theta_l=\frac{g\bigg(\Delta+\frac{2 a M_v}{g}\bigg(r+\rho^2\frac{\partial\alpha}{\partial M}-\beta\frac{\partial \beta}{\partial M}\bigg)\bigg)}{\rho^2 \sqrt{(1-2\alpha) a^2- (\Delta-2 \alpha (a^2+r^2)+\beta^2)\csc^2\theta}}.
         \label{h15}
 \end{equation}
 Again, to ensure that $\theta_l>0$ outside the horizon, we should have
 \begin{equation}
     \Delta+\frac{2 a M_v}{g}\bigg(r+\rho^2\frac{\partial\alpha}{\partial M}-\beta\frac{\partial \beta}{\partial M}\bigg)>0,
 \end{equation}
 as $g>0$. We will analyse the two cases of $\Delta>0$ (for the possibility of horizon lying outside $M+ \sqrt{M^2-a^2}$) and $\Delta<0$ (for the possibility of horizon lying inside $M+ \sqrt{M^2-a^2}$) individually.
 
 First let us consider the case of $\Delta>0$ at the horizon. To achieve this, that is, to get $\theta_l=0$, we should have from Eq.\eqref{h15}
 \begin{equation}
     \begin{split}
         & r+\rho^2\frac{\partial\alpha}{\partial M}-\beta\frac{\partial \beta}{\partial M}>0 \qquad \text{for} \qquad M_v<0 \qquad \text{and} \qquad g>0, \\
         & r+\rho^2\frac{\partial\alpha}{\partial M}-\beta\frac{\partial \beta}{\partial M}<0 \qquad \text{for} \qquad M_v>0 \qquad \text{and} \qquad g>0.
     \end{split}
 \end{equation}
Without risk of losing the generality of the the result, we will assume $\beta=0$ for the rest of the analysis. This gives \begin{equation}
    \begin{split}
        r+\rho^2\frac{\partial\alpha}{\partial M}>0 \qquad \text{for} \qquad M_v<0,\\
        r+\rho^2\frac{\partial\alpha}{\partial M}<0 \qquad \text{for} \qquad M_v>0.
    \end{split}
\end{equation}The lower/upper bound of this inequality gives $\alpha=\frac{\Delta}{2\rho^2}$ with the choice of $(a^2+r^2)/2$ as an integration constant and this is the null vector given in Eq.\eqref{o8} to obtain $\Delta=0$ as the horizon. In general, the solution for $\theta_l=0$ from Eq.\eqref{h15} is given as
 \begin{equation}
     \Delta+\frac{2 a M_v}{g}\bigg(r+\rho^2\frac{\partial\alpha}{\partial M}\bigg)=0,
 \end{equation}
 or
 \begin{equation}
    \bigg(1+\frac{2 a M_v}{g}\frac{\partial\alpha}{\partial M}\bigg)r^2-2\bigg(M-\frac{a M_v}{g}\bigg)r+a^2\bigg(1+\frac{2 a M_v}{g}\frac{\partial\alpha}{\partial M}\cos^2\theta\bigg)=0.
 \end{equation}
 The solution of this equation will be diverging as $\frac{\partial\alpha}{\partial M} \to -\frac{g}{2 a M_v}$, which is the reasonable possibility. We will here give an example to support this argument. But, let us first assume $M_{v v} \approx 0$ (only for this example), which is the reasonable approximation in the semi-classical limit. We then take
 \begin{equation}
     \alpha=\frac{a^2 Q^2+M^2 \csc^2\theta}{2 a^2 Q^2},
 \end{equation}
where $Q=Q(v)$ is some arbitrary function. With this choice of $\alpha$, we get
\begin{equation}
    \gamma=\frac{M^2-a^2 Q^2 \sin^2\theta+2 \sqrt{M Q^2(\rho^2 M+2 a^2 r Q^2 \sin^2\theta)}}{2 a Q^2}.
\end{equation}
We thus have the pair of null vectors given as
\begin{equation}
    \begin{split}
        l_\mu=& \left(-\frac{a^2 Q^2+M^2 \csc^2\theta}{2 a^2 Q^2}, \; 1, \; 0, \; \frac{M^2-a^2 Q^2 \sin^2\theta+2 \sqrt{M Q^2(\rho^2 M+2 a^2 r Q^2 \sin^2\theta)}}{2 a Q^2}\right),\\
        n_\mu=& (-1, \;0, \; 0, \; a \sin^2\theta).
    \end{split}
\end{equation}
They satisfy the relations given in Eq.\eqref{3} and thus can be used for the calculation of expansion scalars. The value of $\theta_l$ is obtained as:
\begin{equation}
    \theta_l=\frac{1}{\rho^2}\bigg(r-M+\frac{M Q(r M+a^2 Q^2 \sin^2\theta)+ Q M_v (\rho^2 M+ a^2 r Q^2 \sin^2\theta)- \rho^2 M^2 Q_v}{\sqrt{M Q^4(\rho^2 M+ 2 a^2 r Q^2 \sin^2\theta)}}\bigg),
\end{equation}
where $Q_v=\partial Q/\partial v$. The solution of $\theta_l=0$ for an apparent horizon will be convenient when $Q^2 \to M_v^2$, for which case, $Q_v \sim 0$. The solution for $r$ in that regime would be
\begin{equation}
    r=\left|\frac{2 M M_v}{M_v^2-Q^2}\right|.
\end{equation}
We thus have diverging $r=r_g$ as $Q^2 \to M_v^2$.
 
 Now, we consider the case of $\Delta<0$ at the horizon. Then, to ensure that $\theta_l=0$, we should have
 \begin{equation}
     \begin{split}
         r+\rho^2\frac{\partial\alpha}{\partial M}<0 \qquad \text{for} \qquad M_v<0,\\
         r+\rho^2\frac{\partial\alpha}{\partial M}>0 \qquad \text{for} \qquad M_v>0.
     \end{split}
 \end{equation}
 Again, the upper/lower bound of this inequality gives  $\alpha=\frac{\Delta}{2\rho^2}$ with the choice of $(a^2+r^2)/2$ as an integration constant, which is the null vector given in Eq.\eqref{o8}. The solution for $\theta_l=0$ from Eq.\eqref{h15} is given as
 \begin{equation}
     \bigg(1+\frac{2 a M_v}{g}\frac{\partial\alpha}{\partial M}\bigg)r^2-2\bigg(M-\frac{a M_v}{g}\bigg)r+a^2\bigg(1+\frac{2 a M_v}{g}\frac{\partial\alpha}{\partial M}\cos^2\theta\bigg)=0.
 \end{equation}
Both for $M_v<0$ and $M_v>0$, proper choice of the value of $\partial\alpha/\partial M$ could make the value of $r=r_g$ recede up to the singularity. This can be seen by minimizing $r$ with respect to $\partial\alpha/\partial M$ in this equation, whose solution is $r^2+a^2\cos^2\theta=0$. Thus, depending on the choice of null vectors (which in fact determines the choice of the foliation), the value of the MOTS can extend from the singularity to the infinity. As we have seen in the previous two examples, all the MOTSs do not foliate an apparent horizon. However, it is not clear which MOTSs dynamically evolve to foliate the apparent horizon and if the apparent horizon is unique\cite[]{19,14}. Moreover, to find the trapping horizon, we have to check each apparent horizon hypersurfaces to confirm whether the space-time inside it is trapped and outside it is normal. So, the problem of locating the trapping horizon extends further beyond the complexity of identifying the apparent horizon.

\subsection{An approximate apparent horizon}\label{2c}

Besides the problem of foliation dependence, there is another more serious problem in general dynamical axially symmetric space-times in the calculation of the trapped surfaces. In general, for time-dependent axial symmetry, it is observed that the future directed outgoing null vector is not tangent to the geodesic everywhere. This is because, the general future-directed outgoing null vector has three independent parameters and can be written as $l_\mu=(-\alpha, \; 1, \; \beta, \; \gamma)$, where $\alpha\ne0$. However, there are four constraints altogether to satisfy by this vector: three independent constraints from the geodesic equation $l^\mu l_{\nu; \mu}=\lambda l_\nu$, for some parameter $\lambda$ (one of the equation $l^\mu l_{r; \mu}=\lambda l_r$ is satisfied identically) and one constraint from the null condition $l^\mu l_\mu=0$. There are thus four equations to satisfy by three variables and this will hold only on some region/hypersurface of the space-time. This is in contrast to the situation in both the time-dependent spherical symmetry and the stationary axial symmetry, where an arbitrary null vector can be made geodesic by the proper choice of a parameter called an affine parameter defining the curve. Thus, an outgoing null vector is a tangent to the geodesic only on some surface in axial symmetry in general and on that surface, the outgoing expansion $\theta_l$ does not vanish usually. A way to address this problem is to calculate an approximate apparent horizon.

To approximate an apparent horizon for the Kerr-Vaidya geometry, let us take the null vectors of Eq.\eqref{o8} for which the solution of $\theta_l=0$ is clearly given by $\Delta=0$. However, as concluded above, the intersection of $v=constant$ and $\Delta=0$ surface does not foliate the apparent horizon in general. The reason behind this is that the vectors of Eq.\eqref{o8} is not tangent to the null geodesics in general. To see this, the geodesic deviation equation $l^\mu l_{\nu; \mu}-\kappa l_\nu$ for some parameter $\kappa$ is given as
\begin{align}
    & l^\mu l_{v; \mu}-\kappa l_v=\frac{a^2 r \sin^2\theta M_v}{\rho^4},\\
    & l^\mu l_{r; \mu}-\kappa l_r=0,\\
    & l^\mu l_{\theta; \mu}-\kappa l_\theta=0,\\
    & l^\mu l_{\phi; \mu}-\kappa l_\phi=-\frac{a r (a^2+r^2) \sin^2\theta M_v}{\rho^4}.
\end{align}
Thus the null vectors are offset from being the geodesics by an order of $M_v$, which is obviously small in the semi-classical limit as pointed out in Sec.\ref{s2e}. So in the semi-classical region, we can approximate $\Delta=0$ as an apparent horizon, and we will explain below that this satisfies all the property for being the slowly evolving horizon. But before that, we will present a technique that is similar to the perturbation expansion to approximate an apparent horizon. Possibly, the validity of this approximation method would be for all $|M_v|<1$ and not only for very small $M_v$.

For this, we proceed forward by making a slight modification on the null vectors of Eq.\eqref{o8}
\begin{equation}
    \begin{split}
        & l_\mu=\bigg(-\frac{\Delta+\lambda(r,\theta) M_v}{2\rho^2},\; 1,\; \zeta(r,\theta) M_v,\; \frac{a \Delta \sin^2\theta+\nu(v,r,\theta)}{2\rho^2}\bigg),\\
        & n_\mu=\bigg(-1,\; 0,\; 0,\; a \sin^2\theta \bigg).
        \label{o25}
    \end{split}
\end{equation}
The null condition $l^\mu l_\mu=0$ gives
\begin{equation}
    \nu=\sin^2\theta \bigg(a\lambda M_v -2a\rho^2+2\rho^2 \sqrt{a^2+\csc^2\theta \lambda M_v-\csc^2\theta \zeta^2 M_v^2}\bigg).
    \label{o44}
\end{equation}
We now calculate the outgoing expansion $\theta_l$ at $r=r_g$ given by the solution of the equation $M=\frac{a^2+r^2+f(t,r,\theta) M_v}{2 r}$ where $f(t,r,\theta)$ is another arbitrary function. Assuming $M_{v v}\sim 0$, we get, from the solution of $\theta_l=0$
\begin{widetext}
\begin{equation}
    f= \frac{2\rho^2 \left(\left(2 \frac{\partial\zeta}{\partial \theta}-\frac{\partial\lambda}{\partial r}\right) \sqrt{\csc ^2 \theta M_v \left(\lambda-M_v \zeta^2\right)+a^2}+2 \zeta \left(\cot \theta \sqrt{\csc ^2 \theta M_v \left(\lambda-M_v \zeta^2\right)+a^2}-a M_v \frac{\partial\zeta}{\partial r}\right)+a \frac{\partial\lambda}{\partial r}\right)}{4 r \sqrt{\csc ^2 \theta M_v \left(\lambda-M_v \zeta^2\right) +a^2}} +\lambda.
    \label{o27}
\end{equation}
\end{widetext}
Again, solving the geodesic deviation equation at $M=\frac{a^2+r^2+f(t,r,\theta) M_v}{2 r}$, with the value of $f$ given by Eq.\eqref{o27}, we get
\begin{widetext}
\begin{align}
    & l^\mu l_{v; \mu}-\kappa l_v=\frac{M_v \left(\left(r^2-a^2\right) \lambda+4 a^2 r^2 \sin ^2\theta \right)}{4 r \rho^4}+{\cal O}\left(M_v^2\right),\\
    & l^\mu l_{r; \mu}-\kappa l_r=0,\\
    & l^\mu l_{\theta; \mu}-\kappa l_\theta =\frac{\left(a^2-r^2\right) M_v \zeta}{2r \rho^2}+{\cal O}\left(M_v^2\right),\\
    & l^\mu l_{\phi; \mu}-\kappa l_\phi= \frac{\left(a^2+r^2\right) M_v \left(\left(a^2-r^2\right) \lambda-4 a^2 r^2 \sin ^2\theta \right)}{4a \rho^4 r}+{\cal O}\left(M_v^2\right),
\end{align}
\end{widetext}
where ${\cal O}\left(M_v^2\right)$ represents the terms of order $M_v^2$ and higher. Now, requiring the geodesic deviation of the null tangent vectors to be of at least ${\cal O}\left(M_v^2\right)$ we should have
\begin{equation}
    \lambda=\frac{4 a^2 r^2 \sin ^2\theta}{a^2-r^2} \qquad \textrm{and} \qquad \zeta=0.
    \label{o50}
\end{equation}
Substituting this in Eq.\eqref{o27}, we get
\begin{widetext}
\begin{equation}
    f=\frac{2 a^2 \sin ^2\theta \left(2 a^4 \cos ^2\theta+2 a^2 r^2-\left(a^4 \cos 2 \theta+a^4+2 r^4\right) \sqrt{\frac{4 r^2 M_v}{a^2-r^2}+1}\right)}{(a-r)^2 (a+r)^2 \sqrt{\frac{4 r^2 M_v}{a^2-r^2}+1}}=\frac{4 a^2 r^2 \sin ^2\theta }{a^2-r^2}+{\cal O}\left(M_v\right).
    \label{o33}
\end{equation}
\end{widetext}
Thus, the solution of the equation $M=\frac{a^2+r^2+f(t,r,\theta) M_v}{2 r}$ with the value of $f$ from Eq.\eqref{o33} gives the value of $r$. This $r$ corresponds to the apparent horizon of the Kerr-Vaidya space-time up to the first order correction in $M_v$. Specifically,
\begin{equation}
    r=M+\sqrt{M^2-a^2}+\frac{a^2 \sin ^2 \theta \left(2 M^2 -a^2+ 2 M \sqrt{M^2-a^2}\right)} {\left(M^2-a^2\right)\left(M+ \sqrt{M^2-a^2}\right)} M_v.
    \label{o52}
\end{equation}
For the Kerr-Vaidya metric in advanced coordinates, when $M_v <0$, the apparent horizon is ellipsoid flattened at the equator, provided that $a$ is small. Similarly, when $M_v >0$, for small $a$, the apparent horizon is ellipsoid flattened at the pole. It can be shown that this approximate horizon is not the unique approximate apparent horizon. There exists other surfaces where the null vectors satisfy geodesic equations up to the second order in $M_v$ and has the vanishing outgoing expansion $\theta_l$ (for example, the exact procedures for the calculation of an approximate horizon applied to the null vectors of Eq.\eqref{s9} give different surface as an approximate horizon). However, these surfaces differs from each other and from the stationary black hole horizon, at most, by an order of $M_v$.

The MOTSs foliating the apparent horizon is unique\cite[]{19}. So, if an apparent horizon is known a priori, then we can infer that this has been foliated by a dynamical evolution of the unique MOTS. This is unlike the case of isolated horizons where foliations are freely deformable. However, we could not say anything about the uniqueness of the apparent horizon itself. (Remark: the assumption of the finite time formation of an apparent horizon constrains its location. However, it is not sufficient constraint to give the unique apparent horizon.)

The approximate apparent horizon calculated here has promising features which we list below and this might provide a clue for the appropriate choice of null vectors for the calculation of trapped surface. The null vectors given in Eq.\eqref{o25} have the vanishing shear tensor in the leading order approximation in $M_v$. The shear tensor $\sigma_{\mu\nu}$ can be calculated by using the relation
\begin{equation}
    \sigma_{\mu\nu}=\tilde{B}_{\mu\nu}-\frac{\theta}{2}h_{\mu\nu},
    \label{d37}
\end{equation}
where $h_{\alpha\beta}=g_{\alpha\beta}+l_\alpha n_\beta + n_\alpha l_\beta$ and $\tilde{B}_{\alpha\beta}=\frac{1}{2}\theta h_{\alpha\beta}$ (see \cite{3} for details). Now, at the apparent horizon, $\theta_l=0$. Direct calculation using Eq.\eqref{d37} gives
\begin{equation}
    \sigma^l_{\mu\nu} \sigma_n^{\mu\nu}= \frac{4 a^2 r_g^2 \cos ^2\theta}{\rho^4 \left(a^2-r_g^2\right)} M_v+ {\cal O} \left(M_v^2\right),
    \label{h59}
\end{equation}
which is already first order in $M_v$. The choice of null vectors with vanishing shear tensor, at least, in the leading order approximation is the natural choice of null vectors (and hence the natural choice of foliation) for the calculation of the trapped surface. This can be explained as follows:

It is undoubtedly true that for the calculation of trapped surfaces in spherically symmetric space-times, we should choose the foliation that respects the symmetry of the space-time\cite[]{33}. For this spherical foliation of choice, the null geodesic congruence normal to this surface is radial and shear free. Some choice of non-spherical foliations, even in stationary spherically symmetric space-times, do not contain trapped surfaces\cite[]{18}. Obviously, choosing an arbitrary axially symmetric foliation does not work for axially symmetric space-times. Shear free geodesic congruence is the axially symmetric analogy of radial geodesic in the spherical symmetric space-time\cite[]{28}. Thus, the foliation associated with the shear free geodesic in axially symmetric space-time compliments the symmetry respecting foliation of the spherically symmetric space-time.

Furthermore, our calculation of an approximate apparent horizon also follows from an assumption of the validity of the geometric horizon conjecture stated in Sec.\ref{1c}. The corollary of the Goldberg and Sachs\cite{34} theorem given in \cite{28} (p.63) implies that if the congruence formed by the two principal null directions $l^\mu$ and $n^\mu$ are geodesic and shear free then the space-time is algebraically special. As the null geodesics of Eq.\eqref{o25} for the calculation of an approximate horizon is shear free, at least in first-order approximation in $M_v$, the congruence formed by them is more algebraically special. This implies that the approximate apparent horizon is also an approximate geometric trapped surface. Moreover, the procedure for the calculation of an approximate apparent horizon is the generalization of the third procedure for the calculation of the geometric horizon, explained above, to the more general case of axial symmetry.

\subsection{Features of the apparent horizon}\label{2d}

We now take the pair of null vectors given in Eq.\eqref{o25} with the value of respective parameters $\nu$ given in Eq.\eqref{o44} and $\lambda$ and $\mu$ given in Eq.\eqref{o50}. Direct calculation using these null vectors yield
\begin{equation}
    \begin{split}
        & {\cal L}_l \theta_l=l^\mu (\theta_l)_{;\mu}=-\frac{2 r^2 M_v}{\rho^4}+{\cal O}\left(M_v^2\right),\\
        & {\cal L}_n \theta_l=n^\mu (\theta_l)_{;\mu}= \frac{a^2-r^2}{\rho^4}-\frac{4 a^2 r^2 \left(a^2+r^2\right) \sin ^2\theta }{\rho^4 \left(a^2-r^2\right)^2}M_v +{\cal O} \left(M_v^2\right).
    \end{split}
\end{equation}
We substitute this in Eq.\eqref{e16} to get:
\begin{equation}
    C=\frac{2 r^2 M_v}{r^2-a^2}+{\cal O} \left(M_v^2\right).
\end{equation}
From this equation, $M_v<0$ gives $C<0$ thereby implying that the apparent horizon of the Kerr-Vaidya line-element in advanced coordinates is timelike and receding in $M_v<0$ domain. These are important features, as $M_v<0$ regime of the Kerr-Vaidya line-element in advanced coordinate is believed to be the evaporating black hole solution of the Einstein equation. Similarly, for $M_v>0$, the apparent horizon is spacelike and advancing.

Again, for the null vectors of Eq.\eqref{o25}, we get $\theta_n=-2 r/\rho^2$. We thus have the condition for the slowly evolving horizon
\begin{equation}
    \sqrt{|C|}\theta_n=\sqrt{\frac{8 r^4 |M_v|}{\rho^4 (r^2-a^2)}}<\frac{\epsilon}{r^2+a^2},
\end{equation}
that holds only when the space-time is slowly rotating, that is, $a\ll r$ and $\sqrt{|M_v|}\sim {\cal O}(\epsilon)$. But, in the semi-classical limit, for slowly rotating case near the apparent horizon $r_g$, $|M_v| \approx r_g'\sim 10^{-3}-10^{-4}$\cite[]{26,27}. This gives $\epsilon \sim 0.01\ll1$. Thus, the apparent horizon of an advanced Kerr-Vaidya line-element, in the semi-classical domain, satisfies the condition to be called as the slowly evolving horizon provided $a\ll r$.

Similarly, the calculation using the pair of null vectors of Eq.\eqref{o8} yields $C \sim M_v$ for $a\ll r$. This again leads to the conclusion that the surface $\Delta=0$ can be called as the slowly evolving horizon when $\sqrt{M_v} \sim \epsilon$. Thus, $\Delta=0$ is the slowly evolving horizon of the Kerr-Vaidya metric in advanced coordinates provided $a\ll r$.

\section{Another possible characterization of a black hole boundary}\label{4}

To begin with, we first summarize the complications we encounter in identifying trapped surfaces as a black hole boundary. Identification of trapped surfaces as black hole boundary is straightforward and poses no problem in stationary space-times (where the event horizon can also be undoubtedly located) and in dynamical spherically symmetric space-times (where the location of the event horizon is problematic). However, trapped surfaces turned out to be the ill-defined concept when we try to explore them beyond these two classes of space-times. This is because of the two fundamental reasons:
\begin{itemize}
    \item The problem of the foliation dependence of trapped surfaces: Even in the simplest class of space-times like Schwarzschild and Vaidya\cite{5}, trapped surface depends on the choice of foliation on which the family of null congruence is orthogonal.
    
    \item An arbitrary outgoing null vector $l_\mu$ is not geodesic everywhere in general axisymmetric space-time.  In general, the region where $l_\mu$ is geodesic is not where the outgoing null expansion $\theta_l$ vanishes. Because of this, even an exact location of a geometrical trapped surface which is a foliation independent entity is not possible.
\end{itemize}

In an attempt to address these problems, we present an alternative way of characterization of a black hole boundary. The common procedure for locating a black hole boundary is by looking for some identically non-zero invariant quantity whose value vanishes on the horizon. For example, the expansion scalar of an outgoing null geodesic vanishes at the trapped surface. Similarly, the norm of the Killing vector field or some particular combinations of the scalar polynomial curvature invariants or Cartan invariants vanishes at the geometrical horizon. So, analogously, we are also looking for an invariant quantity that is identically zero at the horizon and non-zero elsewhere. However, for us, that quantity should be foliation/observer-independent and applicable to the more general case of dynamical axial symmetry. Such a quantity would be an arbitrary radial trajectory $l^r$, which becomes null on the horizon, that is, $l^\mu l_\mu=0$, given $l^\theta$ and $l^\phi$ are null (which can be constructed by choice).

To demonstrate that all the radial trajectory indeed becomes null at the horizon, we here given an example of the general spherically symmetric space-time
\begin{equation}
    d\tau^2= - e^{2 h(t,r)} f(t,r) dt^2+\frac{1}{f(t,r)} dr^2+ r^2 d\Omega^2.
    \label{s47}
\end{equation}
Now, for this space-time, the equation of motion in $\theta$ direction is
\begin{equation}
    \frac{d}{d\tau}\frac{\partial {\mathcal L}}{\partial \dot\theta}=\frac{d}{d\tau}(r^2 \dot \theta)=-\frac{\partial {\mathcal L}}{\partial \theta}= r^2 \sin\theta \cos\theta \left(\frac{d\phi}{d\tau}\right)^2,
\end{equation}
where $\dot \theta=d\theta/d\tau$. So, if we initially choose $\theta=\pi/2$ and $\dot \theta=0$ then, $\ddot\theta=0$. This implies that the geodesic motion can be described in an invariant plane and we choose that plane to be the equator $\theta=\pi/2$. Thus, the radial equation of motion for the space-time of Eq.\eqref{s47} can be written as
\begin{equation}
    \dot r^2= -\delta f(t,r)+ \left(e^{h(t,r)} f(t,r) \dot t\right)^2-\frac{L^2}{r^2}f(t,r),
\end{equation}
where $\delta=0$ for null geodesic, $\delta=1$ for timelike geodesic and $L=r^2 \frac{d\phi}{d\tau}$ is constant. As $\dot t$ does not depend on the geodesic being null or timelike and as $L$ is arbitrary, the only surface where the radial trajectory $\dot r$ is always null is $f(t,r)=0$. So, in dynamical spherically symmetric space-times, all the radial trajectory becomes null on some hypersurface, and this three-surface is uniquely characterized as the black hole boundary by our prescription. Also, looking at the geodesic equations for the Kerr space-time given in \cite{28}, we can see that all of the timelike radial geodesics approach null geodesic on $r^2+a^2- 2r M =0$ surface. This surface is, in fact, both the event and the apparent horizon of the Kerr space-time.

Hence, it might be reasonable to characterize a black hole boundary as the hypersurface where every timelike radial geodesics approaches null geodesics. This is the asymptotic three surface, and the motivation for this is the classical picture of the black hole as an asymptotic state of the gravitational collapse. For most of the space-times, the black hole boundary in this characterization is given by the solution of $g^{rr}=0$. Some space-times, like Vaidya, has $g^{rr}=0$ identically and non-zero $g^{t r}$. Coordinate transformation is possible for such space-times to make $g^{tr}=0$, and then the solution of $g^{rr}=0$ gives the black hole boundary (otherwise, the full geodesic equation should be solved).

We now take an example of the general axisymmetric space-time with two parameters
\begin{equation}
    d\tau^2=e^{2 h}\bigg(1-\frac{2 M r}{\rho^2}\bigg)dt^2+\frac{4 e^h a M r \sin^2\theta}{\rho^2}dt d\phi
    -\frac{\rho^2}{\Delta}dr^2
    -\rho^2 d\theta^2-\frac{(r^2+a^2)^2-a^2 \Delta \sin^2\theta}{\rho^2}\sin^2\theta d\phi^2,
    \label{n69}
\end{equation}
where $h=h(t,r,\theta)$ and $M=M(t,r,\theta)$. To calculate the boundary of the black hole for this space-time, we assume the trajectory of the form
\begin{equation}
    l^\mu =\beta \left(1,\; \alpha, \; \gamma, \; \delta \right),
\end{equation}
where $\beta$, $\alpha$, $\gamma$ and $\delta$ are the functions of $t$, $r$ and $\theta$. Now, assuming $l^\mu$ to be a time like trajectory, it should satisfy $l^\mu l_\mu =-1$ and this gives
\begin{equation}
    \alpha_t=\sqrt{\frac{\Delta}{\rho^2}\left(-\frac{1}{\beta^2}- \frac{2 r M \left(e^h -a\sin^2\theta \delta \right)^2}{\rho^2}+ e^{2h} -\rho^2\gamma^2- \left(a^2+ r^2 \right)\sin^2\theta \delta^2\right)}.
    \label{r60}
\end{equation}
Now, this time like radial trajectory coincides identically with the null trajectory given by
\begin{equation}
    \alpha_n= \sqrt{\frac{\Delta}{\rho^2}\left(- \frac{2 r M \left(e^h -a\sin^2\theta \delta \right)^2}{\rho^2}+ e^{2h} -\rho^2\gamma^2- \left(a^2+ r^2 \right)\sin^2\theta \delta^2\right)},
\end{equation}
on the surface $\Delta=0$ irrespective of the form of the parameters $e^h$, $\gamma$ and $\delta$ assuming that the normalization factor $\beta \ne 0$ there (this is expected otherwise, the trajectory is identically zero at $\Delta=0$. Thus, for the general form of the axisymmetric metric given by Eq.\eqref{n69}, adopting the definition presented here gives $g^{rr}=0$ as the black hole boundary.

\section{Conclusion and Discussions}

It seems that the quasi-local measures for determining a black hole boundary are not suitable for the general axisymmetric space-times. In spherical symmetry, there exists a preferred choice of foliation which obeys the symmetry of the space-time and gives a unique MOTS. However, such a preferred choice of foliations can not be made in axisymmetric space-times. We have shown that for axisymmetric space-times, MOTS is not unique and lies anywhere from the singularity to the infinity depending on the choice of the null vectors. Although all these MOTSs do not foliate apparent horizons, many of them could. However, there is not a preferred choice of MOTS foliating an apparent horizon that acts as a true black hole boundary.

Another problem in the calculation of an apparent horizon as explained in Sec.\ref{2b} is that the outgoing null vectors are not geodesic everywhere in general in axisymmetric space-times. At least, for Kerr-Vaidya line-element, it is found that the surface where $\theta_l=0$ does not coincide with the hypersurface where the null vectors are geodesic. We expect the same situation to occur in most of the general axisymmetric space-times. If this is the case, then the calculation of the trapped surface as a black hole boundary does not even make a sense for axisymmetric space-times. One way to get out of this problem was the calculation of an approximate apparent horizon, explained in Sec.\ref{2c}. The approximate horizon exists only for small $M_v$, and when it exists, Sec.\ref{2d} shows that it has some promising features. The problem, however, is that the approximate apparent horizon calculated in the way given in Sec.\ref{2c} is not unique.

We have obtained an approximate apparent horizon that has some promising features. This encouraging result has been obtained by using the null vectors of Eq.\eqref{o25} that has the vanishing shear tensor $\sigma_{\mu\nu}$ in the leading order. The null congruence used for the calculation of an approximate horizon being more algebraically special makes it eligible to be called as the geometric trapped surface. The null vectors with vanishing shear tensor $\sigma_{\mu\nu}$, at least in the leading order, could thus be a natural choice for calculation of the trapped surface. The choice of null vectors determines the choice of foliation.

We have also proposed a new way to characterize a black hole boundary. Although our procedure for locating the black hole boundary requires solving the full radial geodesic equation of the space-time in general, for some forms of space-time it is just the solution of $g^{rr}=0$. Eq.\eqref{n69} is an example of such types of space-time. The validity of our approach for locating the black hole boundary lies in the fact that such a surface exists in the black hole solution of the Einstein equation (for example, the Schwarzschild and the Kerr solution). The universality of our approach relies on whether such surface where all the timelike radial geodesics approaching the null geodesics exists in all black hole solutions. From our study on the black hole boundary, we have noticed that the definition of the black hole boundary plays a crucial role in our understanding of black holes. The new definition is therefore exciting and is believed to meet this expectation.

\begin{acknowledgments}
Pravin Kumar Dahal is supported by IMQRES.
\end{acknowledgments}

\nocite{*}



\end{document}